\documentclass[12pt]{article}
\usepackage[english]{babel}
\usepackage{amsmath}
\usepackage{amssymb}
\usepackage[pdftex]{color}
\usepackage{graphicx}
\usepackage[mathscr]{eucal}%
\usepackage{amsfonts}%
\usepackage{cite}
\begin{document}

\small{\title{\bf Nonlinear Operator Superalgebras and BFV--BRST
Operators for Lagrangian Description of Mixed-\-sym\-met\-ry HS
Fields in AdS Spaces}}

\author{{A.A. Reshetnyak}
\\
\textit{Institute of Strength Physics and Materials Science}\\
\textit{of SB of Russian Academy of Sciences, 634021 Tomsk,
Russia}}
\date{}

\maketitle \thispagestyle{empty}
\renewcommand{\abstractname}{}

\begin{abstract}
 {\small We study the properties of nonlinear
 superalgebras $\mathcal{A}$ and algebras $\mathcal{A}_b$ arising
 from a one-to-one correspondence between the sets of
 relations that extract AdS-group irreducible representations
 $D(E_0,s_1,s_2)$ in AdS$_d$-spaces and the sets of operators that
 form $\mathcal{A}$ and $\mathcal{A}_b$, respectively, for fermionic,
 $s_i=n_i+\frac{1}{2}$, and bosonic, $s_i=n_i$,  $n_i \in \mathbb{N}_0$,
 $i=1,2$, HS fields characterized by a Young tableaux with two rows.
 We consider a method of
 constructing the Verma modules $V_\mathcal{A}$, $V_{\mathcal{A}_b}$
 for $\mathcal{A}$, $\mathcal{A}_b$ and establish a possibility of
 their Fock-space realizations in terms of formal power
 series in oscillator operators which serve to realize
 an additive conversion of the above (super)algebra
 ($\mathcal{A}$) $\mathcal{A}_b$, containing a set of
 2nd-class constraints, into a converted (super)algebra
 $\mathcal{A}_{b{}c}$=$\mathcal{A}_{b}$+$\mathcal{A}'_b$
 ($\mathcal{A}_c$=$\mathcal{A}$+$\mathcal{A}'$), containing
 a set of 1st-class constraints only. For the algebra
 $\mathcal{A}_{b{}c}$, we construct an exact nilpotent
 BFV--BRST operator $Q'$ having nonvanishing terms of 3rd degree
 in the powers of ghost coordinates and use $Q'$ to construct
 a gauge-invariant Lagrangian formulation (LF) for HS fields with
 a given mass $m$ (energy $E_0(m)$) and generalized spin
 $\mathbf{s}$=$(s_1,s_2)$. LFs with off-shell algebraic constraints
 are examined as well.
}
\end{abstract}

 \thispagestyle{empty}
\renewcommand{\refname}{\center\underline{\phantom{ReferencesReference}}}
\renewcommand{\thesubsubsection}{\arabic{subsubsection}.}

\vspace{1ex} {\center\subsubsection{ INTRODUCTION}\label{int}}
\vspace{1ex}

The growing interest in field-theoretical models of higher
dimensions is due to the problem of a unified description of the
known interactions and the variety of elementary particles, which
becomes especially prominent at high energies (partially
accessible to the LHC), thus stimulating the present-day
development of a mixed-symmetry higher-spin (HS) field theory in
view of its close relation to superstring theory in
constant-curvature spaces \cite{Beisert, Heslop}, which operates
with an infinite set of bosonic and fermionic HS fields (that
correspond to arbitrary tensor representations of the Wigner
little algebra) subject to a multi-row Young tableaux (YT)
$Y(s_1,...,s_k)$, $k \geq 1$, whose study has been initiated by
\cite{Curtright} and continued in \cite{mixsymmLabastida}; for a
review on HS field theory, see \cite{reviews1, reviews2}. The
theory of free and interacting mixed-symmetry HS fields has been
developed in the framework of various approaches, which may be
classified as the light-cone formalism \cite{Metsaev2}, Vasiliev's
frame-like formalism \cite{Vasiliev, Alkalaev, Grigoriev} using
the unfolded approach \cite{unfolded}, and Fronsdal's
\cite{Fronsdal}, both constrained \cite{ZinovievbosAdS0211233} and
unconstrained \cite{Sagnotti}, metric-like  formalism. While the
results of constructing a Lagrangian formulation (LF) for free
bosonic mixed-symmetry HS fields in the flat space are well-known
within all of these approaches, see for instance \cite{Bekaert,
flatfermmix, 0707.2181, Skvortsov}, the corresponding results for
the AdS$_d$-space have been developed in the light-cone, for
the $AdS_5$-space \cite{Metsaev}, and in the frame-like, for an integer spin
\cite{Vasiliev3, ZinovievbosAdS0809.3287}, formulations and remain
unknown in a more involved case of half-integer spins with
a YT $Y(s_1,...,s_k)$, $s_i = n_i+\frac{1}{2}, n_i \in \mathbb{N}_0$.

The present article is devoted to solving this problem for free
integer and half-integer HS fields in the AdS$_d$-space that are
subject to a YT with two rows, in unconstrained and constrained
metric-like formulations, on a basis of the (initially elaborated for
a Hamiltonian quantization of gauge theories, and being universal
for all of the above constructions) BFV--BRST formalism
\cite{BFV,BFV2}; see the review \cite{Henneaux} as well.
The basic idea here consists in a solution of a problem
\textit{inverse} to that of the method \cite{BFV}, just as in
string field theory \cite{SFT}, in the sense of constructing
a gauge LF with respect to a nilpotent BFV--BRST operator $Q$,
constructed, in turn, from a system $\{O_\alpha\}$ of 1st-class
constraints that include a special nonlinear \textit{non-gauge}
operator symmetry (super)algebra
$(\mathcal{A}_c)\mathcal{A}_{bc}$ for (half-)integer HS fields
$\{O_I\}$: $\{O_I\} \supset \{O_\alpha\}$. These quantities
$\{O_I\}$ correspond to the initial AdS$_d$-group irreducible
representation (irrep) relations extracting the spin-tensors of a
definite mass $m$ (including $m=0$) and spin (except for such
algebraic conditions as the gamma- and traceless conditions
in the case of a constrained description) and realized as operator
constraints for a vector of a special Fock space whose coefficients
are (spin-)tensors related to the spin of the basic HS field. As a
result, the final action and the sequence of reducible gauge
transformations are reproduced by means of the simplest operations
of decomposing the resulting gauge vectors of the Hilbert space
that contain the initial HS (spin-)tensors and the gauge parameters
with respect to the initial oscillator and ghost variables,
subject to the spin and ghost number conditions, and also by means
of calculating the corresponding scalar products, first realized
in \cite{Ouvry, Pashnev}. Due to the required presence of
auxiliary (spin-)\linebreak tensors with a lesser spin, in order to have a
closed LF for the basic (spin-)tensor with a given spin (mentioned
in the pioneering works of Fierz--Pauli \cite{FP} and Singh--Hagen
\cite{SH} as a crucial part in the definition of a correct number
of physical degrees of freedom), there arises a necessity of converting
the sub(super)algebra of the total HS symmetry (super)algebra
corresponding to the subset of 2nd-class constraints into that of
1st-class constraints. This conversion procedure is realized as
an additive version of \cite{conversion, conversion1}, by means
of constructing the Verma modules \cite{Dixmier} for Lie (super)algebras
corresponding to HS fields in the flat case and for specially
deformed nonlinear (super)algebras $(\mathcal{A}')\mathcal{A}_b'$
in the AdS$_d$-space for HS fields with $Y(s_1)$; see
\cite{0206027, adsfermBKR}. A transition to mixed-symmetry HS
fields with $Y(s_1,...,s_k), k\geq 2$ meets a significant
obstacle to an application of a Cartan-like decomposition for
$(\mathcal{A}')\mathcal{A}'_b$, which is one of the goals of the
present article.

Another aspect concerns the structure of the BFV--BRST operator $Q$,
being more involved in the case of a converted nonlinear (super)algebra for
HS fields subject to $Y(s_1,s_2)$, $(\mathcal{A}_{c})\mathcal{A}_{b{}c}$,
in view of the presence of nonvanishing terms of 3rd order in the powers
of ghosts, because of a nontrivial character of the Jacobi identity for $O_I$,
in comparison, first, with $Q$ for (half-)integer HS fields in the flat space
\cite{flatfermmix,Pashnev}, second, with $Q$ for totally-symmetric
(half-)integer HS fields in the $AdS_d$-space \cite{adsfermBKR,0206027},
and, third, with $Q$ for special classical quadratic (super)algebras investigated
in \cite{Sevrin, 0701243}, because of a partially nonsupercommuting character
of the operators $O_I$.

The paper is organized as follows. In Section~2, we examine the
initial operator (super)algebra $(\mathcal{A})\mathcal{A}_{b}$. In
Section~3, we consider Proposition, which determines a way to
obtain algebraic relations for the (super)algebras of the parts
$(\mathcal{A}')\mathcal{A}'_{b}$ additional to those for a
specially modified (super)algebra
$(\mathcal{A}_{mod})\mathcal{A}_{b{}mod}$, and examine a
construction of Verma modules that realize the highest-weight
representation of $(\mathcal{A}')\mathcal{A}'_{b}$ and their
realization in an auxiliary Fock space. An exact BFV--BRST
operator for a converted (super)algebra
$(\mathcal{A}_{c})\mathcal{A}_{b{}c}$ is obtained in Section~4, on
the basis of a solution of the Jacobi identity, due to the absence
of non-trivial higher-order relations for
$(\mathcal{A}_{c})\mathcal{A}_{b{}c}$. The action and the sequence
of reducible gauge transformations, mainly for bosonic HS fields
of a fixed spin $\mathbf{s}=(s_1,s_2)$, are deduced in Section~5.
In the conclusion, we summarize the results of this article and
discuss some open problems.

We mainly use the conventions of
Refs.~\cite{flatfermmix,adsfermBKR}.

 \vspace{2ex} {\center\subsubsection{NONLINEAR
(SUPER)ALGEBRA FOR MIXED-SYMMETRY HS FIELDS IN ADS
SPACE-TIME}}\label{nlnsuperalg} \vspace{1ex}

A massive spin $\mathbf{s}=(s_1,s_2)$, $s_i = n_i+\frac{1}{2}$,
$n_1 \geq n_2$, representation of the AdS group in an AdS$_d$ space
is realized in a space of mixed-symmetry spin-tensors with a suppressed
Dirac index, and is characterized by $Y(s_1,s_2)$,
\begin{equation}\label{Young k2}
\Phi_{(\mu)_{n_1},(\nu)_{n_2}} \hspace{-0.2em}\equiv
\hspace{-0.2em}
\Phi_{\mu_1\ldots\mu_{n_1},\nu_1\ldots\nu_{n_2}}(x)
\hspace{-0.3em}\longleftrightarrow \hspace{-0.3em}
\begin{array}{|c|c|c c c|c|c|c|c|c| c| c|}\hline
  \!\mu_1 \!&\! \mu_2\! & \cdot \ & \cdot \ & \cdot \ & \cdot\ & \cdot\ & \cdot\  & \cdot\ &
  \cdot\
  & \cdot\    &\!\! \mu_{n_1}\!\! \\
   \hline
    \! \nu_1\! &\! \nu_2\! & \cdot\
   & \cdot\ & \cdot & \cdot & \cdot & \cdot & \cdot & \!\!\nu_{n_2}\!\!   \\
  \cline{1-10}
\end{array}\ ,
\end{equation}
subject to the following equations ($\beta = (2;3) \Longleftrightarrow
(n_1>n_2; n_1 = n_2)$; $r$ being the inverse squared AdS$_d$
radius, and Dirac's matrices satisfying the relation
$\bigl\{\gamma_\mu,\gamma_\nu\bigr\}=2g_{\mu\nu}(x))$:
\begin{eqnarray}
\label{Eq-0} \Bigl(\bigl[i\gamma^{\mu}\nabla_{\mu}
    -r^\frac{1}{2}(n_1  + \textstyle\frac{d}{2}-\beta)-m
\bigr],\ \gamma^{\mu_1} ,\ \gamma^{\nu_1}
\Bigr)\Phi_{(\mu)_{n_1},\ (\nu)_{n_2}} =
\Phi_{\{(\mu)_{n_1},\nu_1\}\nu_2...\nu_{n_2}}=0.
\end{eqnarray}
For a simultaneous description of all half-integer HS fields, one
introduces a Fock space $\mathcal{H}$, generated by 2 pairs of
creation $a^i_\mu(x)$ and annihilation $a^{j+}_\mu(x)$ operators,
$i,j =1,2, \mu,\nu =0,1...,d-1$: $[a^i_\mu,
a_\nu^{j+}]=-g_{\mu\nu}\delta_{ij}$, and a set of constraints for
an arbitrary string-like vector $|\Phi\rangle \in \mathcal{H}$,
\begin{eqnarray}
\label{t't0}  {\tilde{t}}'_0|\Phi\rangle  &=&
\bigl[-i\tilde{\gamma}^\mu D_\mu + \tilde{\gamma}\bigl( m +
\sqrt{r} (g^1_0 - \beta)\bigr)\bigr]|\Phi\rangle=0 ,\\
\label{tit}  \bigl({t}^i, t \bigr)|\Phi\rangle & = &
\bigl(\tilde{\gamma}^\mu a^i_\mu, a^{1+}_\mu
a^{2\mu}\bigr) |\Phi\rangle=0,\\
\label{PhysState}  |\Phi\rangle & =&
\textstyle\sum_{n_1=0}^{\infty}\sum_{n_2=0}^{n_1}\Phi_{(\mu)_{n_1},(\nu)_{n_2}}(x)\,
a^{+\mu_1}_1\ldots\,a^{+\mu_{n_1}}_1a^{+\nu_1}_2\ldots\,a^{+\nu_{n_2}}_2|0\rangle,
\end{eqnarray}
given in terms of the operator $D_\mu =
\partial_\mu-\omega_\mu^{ab}(x)\bigl(\sum_{i}a_{i{}a}^+a_{i{}b}
-\frac{1}{8}\tilde{\gamma}_{[a}\tilde{\gamma}_{b]}\bigr),
a^{(+)\mu}(x)=e^\mu_a(x)a^{(+)a},$ equivalent to the covariant
derivative $\nabla_{\mu}$ in its action in $\mathcal{H}$, with a
vielbein $e^\mu_a$, a spin connection $\omega_\mu^{ab}$, and
tangent indices $a,b=0,1...d-1$. The scalar fermionic operators
${\tilde{t}}'_0, {t}^i$ are defined with the help of an extended
set $\tilde{\gamma}^\mu,\tilde{\gamma}$ of Grassmann-odd
gamma-matrix-like objects \cite{adsfermBKR},
$\{\tilde{\gamma}^\mu,\tilde{\gamma}^\nu\} =2g^{\mu\nu}$,
$\{\tilde{\gamma}^\mu,\tilde{\gamma}\}=0$, $\tilde{\gamma}^2=-1$,
related to the conventional gamma-matrices by an odd
non-degenerate transformation: $\gamma^{\mu} =
\tilde{\gamma}^{\mu} \tilde{\gamma}$. The validity of relations
(\ref{t't0}), (\ref{tit}) is equivalent to a simultaneous
fulfilment of Eqs.~(\ref{Eq-0}) for all the spin-tensors
$\Phi_{(\mu)_{n_1},(\nu)_{n_2}}$.

The construction of a Hermitian BFV--BRST charge $Q$, whose
special cohomology in the zero-ghost-number subspace of a total
Hilbert space $\mathcal{H}_{\mathrm{tot}} = \mathcal{H}\otimes
\mathcal{H}'\otimes \mathcal{H}_{\mathrm{gh}}$ will coincide with
the space of solutions of Eqs. (\ref{Eq-0}), implies constructing
a set of 1st-class quantities $O_I$, $\{O_\alpha\}\subset
\{O_I\}$, closed under the operations of \textbf{a)} Hermitian
conjugation with respect to an odd scalar product,
$\langle{\Psi}|\Phi\rangle_1\equiv\langle\tilde{\Psi}|\Phi\rangle$
\cite{flatfermmix}, with a measure $d^dx \sqrt{-{\mathrm{det}}g}$,
and \textbf{b)} supercommutator multiplication $[\ ,\ \}$. As a
result, the final massive (massless for $m=0$) half-integer HS
symmetry superalgebra in a space $AdS_d$ with $Y(s_1,s_2)$,
$\mathcal{A}$ = $\{o_I\}$ = $\{{\tilde{t}}'_0, {t}^i, {t}^{i+}, t,
t^+, {l}^i, {l}^{i+}, {l}_{ij}, {l}_{ij}^+, g_0^i,
\tilde{l}_0'\}$, $i\leq j; i,j=1,2$,
\begin{equation}
 \bigl(t^{i+};g^i_0; t^+;l^i, l^{+i};l_{ij}\bigr) \hspace{-0.2ex}=\hspace{-0.2ex}
  \bigl(\tilde{\gamma}^\mu a_\mu^{i+};
-a^{i+}_\mu a^{\mu{}i}+{\textstyle\frac{d}{2}}; a^{\mu{}1}
a_\mu^{2+};-i(a^{\mu{}i}, a^{+\mu{}i})D_\mu;
{\textstyle\frac{1}{2}}a^\mu_i a_{\mu{}j}\bigr) , \label{l2+}
\end{equation}
\vspace{-2em}
\begin{eqnarray}
{\tilde{l}}'_0=g^{\mu\nu}(D_\nu
D_\mu-\Gamma^\sigma_{\mu\nu}D_\sigma)
-r\Bigl(\sum_i(g^i_0+t^{i+}t^i)+{\textstyle\frac{d(d-5)}{4}}\Bigr)
+ \bigl(m + \sqrt{r} (g^1_0 - \beta)\bigr)^2, \label{l'l0}
\end{eqnarray}
will contain a central charge $\tilde{m}=(m-\beta\sqrt{r})$, a subset
of (4+12) differential $\{l_i, l_i^+\} \subset \{o_\mathbf{a}\}$
and algebraic $ \{t_i,t^+_i, t, t^+, l_{ij}, l_{ij}^+\} \subset
\{o_\mathbf{a}\} $ 2-class constraints, as well as some
particle-number operators $g_0^i$, composing, along with $\tilde{m}^2$,
an invertible supermatrix $\|[o_\mathbf{a}, o_\mathbf{b}\}\| =
\|\Delta_{\mathbf{ab}}(g_0^i,\tilde{m})\| +
\mathcal{O}(o_\mathbf{a})$, and obeys some non-linear
algebraic relations w.r.t. $[\ ,\ \}$. To construct an appropriate
LF, it is sufficient to have a simpler (so-called \textit{modified})
superalgebra $\mathcal{A}_{mod}$, obtained from $\mathcal{A}$ by
a linear nondegenerate transformation of $o_I$ to another basis
$\tilde{o}_I$, $\tilde{o}_I= u^J_Io_J$, $\tilde{\gamma} \notin \{\tilde{o}_I\}$,
so that the AdS-mass term $m_D = (m + \sqrt{r} (n_1 + \frac{d}{2} - \beta))$
factors out of ${\tilde{t}}'_0, {\tilde{l}}'_0$, which change only
to ${{t}}_0 = -i\tilde{\gamma}^\mu D_\mu$, $l_0 = -t_0^2$.

As a result, the operators $\tilde{o}_I$, given by (\ref{tit}),
(\ref{l2+}), with  the central charge $\tilde{m}$, satisfy the
relations given by Table~\ref{til-table},
\begin{table}[t] {\footnotesize
\begin{eqnarray*}\hspace{-1.0em}
\begin{array}{||c||c|c|c|c|c|c|c|c|c|c||c||}\hline\hline\vphantom{\biggm|}\hspace{-0.3em}
[\,\downarrow\,,\to\}\hspace{-0.5em}&
\hspace{-0.5em}t_0\hspace{-0.5em}&t^i&t^{i{}+}& t & t^+ & l_0 &
l^i &l^{i{}+} & l^{ij}
&l^{ij{}+} &g^i_0 \\
\hline\hline\vphantom{\biggm|} t_0
   &\hspace{-0.3em} -2l_0\hspace{-0.3em} & 2l^i & 2l^{i+} &
   0 & 0 & 0 & \hspace{-0.3em}{M}^i\hspace{-0.3em}& \hspace{-0.3em}-{M}^{i+}\hspace{-0.3em}
    & 0 & 0 &0\\
\hline\vphantom{\biggm|} t^k
   &\hspace{-0.3em} 2l^{k}\hspace{-0.3em} &\hspace{-0.3em}4l^{ki}\hspace{-0.3em}&
   \hspace{-0.3em}A^{ki}\hspace{-0.3em} &\hspace{-0.3em}-t^2\delta^{k1}\hspace{-0.3em}
   & \hspace{-0.3em}-t^1\delta^{k2}\hspace{-0.3em} &
   \hspace{-0.3em}2{M}^k \hspace{-0.3em}&
   0 &\hspace{-0.3em} -t_0\delta^{ik}\hspace{-0.3em} & 0 &\hspace{-0.3em} B^{k,ij}\hspace{-0.3em} & \hspace{-0.4em} t^i \delta^{ki}\hspace{-0.4em} \\
\hline\vphantom{\biggm|} t^{k{}+}
   &\hspace{-0.3em} 2l^{k+}\hspace{-0.3em} & \hspace{-0.3em}A^{ik}\hspace{-0.3em} &\hspace{-0.3em}4l^{ki+} \hspace{-0.3em}&
   \hspace{-0.3em} t^{1+}\delta^{k2}\hspace{-0.3em}&  \hspace{-0.3em}
   t^{2+}\delta^{k1}\hspace{-0.3em}
   &\hspace{-0.3em}-2{M}^{k+}\hspace{-0.3em} &
 \hspace{-0.3em}  t_0\delta^{ik}\hspace{-0.3em}
    & 0 & \hspace{-0.3em}- B^{k,ij+}\hspace{-0.3em} & 0 & \hspace{-0.4em}- t^{i+}\delta^{ki} \hspace{-0.4em}\\
\hline\vphantom{\biggm|} t
   & \hspace{-0.3em}0 \hspace{-0.3em}&\hspace{-0.3em} t^2\delta^{i1} \hspace{-0.3em} & \hspace{-0.3em}- t^{1+}\delta^{i2} \hspace{-0.3em}
    & 0 & \hspace{-0.3em}g_0^1 - g_0^2 \hspace{-0.3em}  &
    0 &
   \hspace{-0.3em} l^2\delta^{i1} \hspace{-0.3em}&\hspace{-0.3em} - l^{1+}\delta^{i2} \hspace{-0.3em}&
   \hspace{-0.3em} D^{ij}\hspace{-0.3em} & \hspace{-0.3em} -G^{ij+}\hspace{-0.3em} & \hspace{-0.3em}F^i\hspace{-0.3em} \\
\hline\vphantom{\biggm|} t^+
   &\hspace{-0.3em} 0 \hspace{-0.3em}&\hspace{-0.3em} t^1\delta^{i2}
   \hspace{-0.3em} & \hspace{-0.3em}- t^{2+}\delta^{i1} \hspace{-0.3em}
   & \hspace{-0.3em} g_0^2 - g_0^1 \hspace{-0.3em}
    & 0  & 0 & \hspace{-0.3em}l^1\delta^{i2}\hspace{-0.3em} & \hspace{-0.3em}-l^{2+}\delta^{i1}\hspace{-0.3em}
     &\hspace{-0.3em} G^{ij}\hspace{-0.3em} & \hspace{-0.3em} -D^{ij+}\hspace{-0.3em} &\hspace{-0.3em} -F^{i+} \hspace{-0.3em}\\
\hline\vphantom{\biggm|} l_0
   &\hspace{-0.3em}0\hspace{-0.3em}&\hspace{-0.3em}-2{M}^i
   \hspace{-0.3em}& \hspace{-0.3em}2{M}^{i+}\hspace{-0.3em} & 0 & 0 & 0
   & \hspace{-0.3em}
    -r{\mathcal{K}}^{bi+}_1\hspace{-0.3em} & \hspace{-0.3em}r{\mathcal{K}}^{bi}_1\hspace{-0.3em}
    & 0 & 0 & 0 \\
\hline\vphantom{\biggm|} l^k
   &\hspace{-0.3em}- {M}^k\hspace{-0.3em} &0&\hspace{-0.3em} -t_0\delta^{ik}
   \hspace{-0.3em}& \hspace{-0.3em}- l^2\delta^{k1} \hspace{-0.3em} & \hspace{-0.3em}-l^1\delta^{k2} \hspace{-0.3em}&\hspace{-0.3em}
   r{\mathcal{K}}^{bk+}_1\hspace{-0.3em}
   & \hspace{-0.3em}{W}^{ki} \hspace{-0.3em} & \hspace{-0.3em}{X}^{ki}\hspace{-0.3em}
    & 0 & \hspace{-0.3em}J^{k,ij}\hspace{-0.3em} & \hspace{-0.4em}l^i\delta^{ik}\hspace{-0.4em} \\
\hline\vphantom{\biggm|} l^{k+} &\hspace{-0.3em}
{M}^{k+}\hspace{-0.3em}
   &\hspace{-0.3em} t_0\delta^{ik} \hspace{-0.3em}& 0 & \hspace{-0.3em}l^{1+}
   \delta^{k2}\hspace{-0.3em} & \hspace{-0.3em}l^{2+}\delta^{k1} \hspace{-0.3em} &
   \hspace{-0.3em}-r{\mathcal{K}}^{bk}_1 \hspace{-0.3em}&\hspace{-0.3em}
   -{X}^{ik}\hspace{-0.3em}
   &\hspace{-0.3em} - {W}^{ki+}\hspace{-0.3em}
    &\hspace{-0.3em} -J^{k,ij+}\hspace{-0.3em} & 0 &\hspace{-0.4em} -l^{i+}\delta^{ik}\hspace{-0.4em}  \\
\hline\vphantom{\biggm|} l^{kl}
   &\hspace{-0.5em} 0 \hspace{-0.5em} & 0 &\hspace{-0.3em}B^{i,kl+}\hspace{-0.3em}&\hspace{-0.3em} -D^{kl}\hspace{-0.3em}
   &\hspace{-0.3em} -G^{kl}\hspace{-0.3em} &  0 & 0 &\hspace{-0.3em}  J^{i,kl+}\hspace{-0.3em} & 0 & \hspace{-0.3em}L^{kl,ij}\hspace{-0.3em} &
   \hspace{-0.4em}l^{i\{k}\delta^{l\}i} \hspace{-0.4em}\\
\hline\vphantom{\biggm|} l^{kl+}
   & 0 &\hspace{-0.3em} -B^{i,kl}\hspace{-0.3em} & 0 & \hspace{-0.3em}
   G^{kl+}\hspace{-0.3em}&\hspace{-0.3em} D^{kl+}\hspace{-0.3em} & 0
   &\hspace{-0.3em}
   -J^{i,kl}\hspace{-0.3em} & 0 &\hspace{-0.3em} - L^{ij,kl}\hspace{-0.3em} & 0 &
   \hspace{-0.4em} -l^+_{i\{k}\delta_{l\}i} \hspace{-0.4em} \\
\hline\hline\vphantom{\biggm|} g_0^k
   & \hspace{-0.3em}0 \hspace{-0.3em}&\hspace{-0.3em}
   -t_k\delta^{ik}\hspace{-0.3em}& \hspace{-0.3em}t^{k+}\delta^{ik}
   \hspace{-0.3em}
    &\hspace{-0.3em} -F^k\hspace{-0.3em} &\hspace{-0.3em}F^{k+}\hspace{-0.3em}
    & 0 &\hspace{-0.3em} -l^{k}\delta^{ik}\hspace{-0.3em}
   &\hspace{-0.3em} l^{k+}\delta^{ik}\hspace{-0.3em} &
   \hspace{-0.3em}-l^{k\{i}\delta^{j\}k}\hspace{-0.3em}
    & \hspace{-0.3em} l^+_{k\{i}\delta_{j\}k}
   \hspace{-0.3em} & 0 \\
   \hline\hline
\end{array}
\end{eqnarray*}}
\caption{The superalgebra of the modified initial
operators.}\label{til-table}
\end{table}
where the quantities $A^{ik}$, $B^{k,ij}$,  $D^{ij}$, $F^i$,
$G^{ij}$,  $J^{k,ij}$,  $L^{kl,ij}$ are defined as follows:
\begin{align}
&  A^{ik} = -2(g_0^i\delta^{ik}-t\delta^{i2}\delta^{k1} -
t^+\delta^{i1}\delta^{k2})\,,\label{Akj}&& D^{ij} = l^{\{i
2}\delta^{j\}1},&&
 G^{ij} = l^{1\{i}\delta^{j\}2}  ,\\
   &J^{k,ij} = -
   \textstyle\frac{1}{2}l^{\{i+}\delta^{j\}k},
   \label{Bkij} && B^{k,ij} = -
\textstyle\frac{1}{2}t^{\{i+}\delta^{j\}k}, && F^i =
t(\delta^{i2}-\delta^{i1}),
 \end{align}
\vspace{-5ex}
\begin{eqnarray}
  && L^{kl,ij} =  - L^{kl,ij+} =  \textstyle\frac{1}{4}\bigl\{\delta^{ik}
\delta^{lj}\bigl[2g_0^k\delta^{kl} + g_0^k + g_0^l\bigr]  -
\delta^{ik}\bigl[t\bigl(\delta^{l2}(\delta^{j1}+\delta^{k1}\delta^{kj}) \\
&& + \delta^{k2}\delta^{j1}\delta^{lk}\bigr)
   + t^+\bigl(1\longleftrightarrow 2\bigr)\bigr]-
\delta^{lj}\bigl[t\bigl(\delta^{k2}(\delta^{i1}+\delta^{l1}\delta^{li})+
\delta^{l2}\delta^{i1}\delta^{kl}\bigr)
    + t^+\bigl(1\longleftrightarrow 2\bigr)\bigr]\bigr\},\nonumber\label{Lklij}
\end{eqnarray}
whereas the nonlinear operators $M^i$, ${W}^{ij}$,
${\mathcal{K}}^{bi}_1$, ${X}^{ij}$ are given by
($\varepsilon^{ij} = - \varepsilon^{ji}$, $ \varepsilon^{12}=1$)
\begin{eqnarray}{}{M}^i &=&r\Bigl(2\textstyle\sum_{k}t^{k+}l^{ki}
+g_0^it^i -{\textstyle\frac{1}{2}}t^i -tt^1\delta^{i2}-
t^+t^2\delta^{i1} \Bigr)   , \label{t0li}
\\{}
{} {W}^{ij} & = &{W}^{ij}_b + \textstyle\frac{r}{4}t^{[j}t^{i]} =
2r\varepsilon^{ij}\left[(g_0^2-g_0^1)l^{12} - t l^{11} + t^+
l^{22}\right] + \textstyle\frac{r}{4}t^{[j}t^{i]}   ,
 \label{lilj}\\
{\mathcal{K}}^{bi}_1& = &\Bigl(4\textstyle\sum_{k}l^{ik+}l^k
+l^{i+}(2g_0^i-1) -2l^{2+}t \delta^{i1} - 2l^{1+}t^+
\delta^{i2}\Bigr)  \label{l'0li+} ,
\\
 {} {X}^{ij}
 \hspace{-0.3em} & \hspace{-0.3em}= \hspace{-0.3em}
 &\hspace{-0.3em} \Bigl\{{{l}}_0
+ r\Bigl(\textstyle\sum_{k}\Bigl(g^k_0+t^{k+}t^k \Bigr)
-\textstyle\frac{5}{2} g_0^i + {g_0^i}^2
+t^+t\Bigl)\Bigl\}\delta^{ij} +
r\Bigl\{\textstyle\frac{1}{2}t^{j+}t^i- 4\sum_{k}
l^{jk+}l^{ik}\nonumber \\
{}& &  - (g_0^1+g_0^2
-\textstyle\frac{3}{2})t\delta^{j1}\delta^{i2} - t^+(g_0^1+g_0^2
-\textstyle\frac{3}{2})\delta^{j2}\delta^{i1} +
(g_0^1-g_0^2)\delta^{j1}\delta^{i1}\Bigl\} \label{l'il'j+}.
\end{eqnarray}
It should be noted that for $r = 0$ the superalgebras
$\mathcal{A}$, $\mathcal{A}_{mod}$ are Lie superalgebras
\cite{flatfermmix}, $\mathcal{A}^{Lie}$,
$\mathcal{A}_{mod}^{Lie}$, which obey the condition (1.2)
mentioned for Lie superalgebras in Ref.~\cite{0802.3781IsKriv} for
$\mathcal{A}_{mod}^{Lie}$ and do not obey it for
$\mathcal{A}^{Lie}$.\footnote{Indeed, there are anticommutators,
$[\tilde{t}'_0, t_k\} = 2l^k -
\tilde{\gamma}r^{\frac{1}{2}}t^1\delta^{k1}$, that violate the
requirement $C^i_{mn} = 0$ if
$\varepsilon(i)+\varepsilon(m)+\varepsilon(n)\neq 0$ for the
Grassmann parities $\varepsilon(i) = \varepsilon(\chi_i)$ of the
quantities $\chi_i$ in Eq.~(1.1): $[\chi_m, \chi_n\}=
C^i_{mn}\chi_m$ in~\cite{0802.3781IsKriv}, because
$C^{[t^1]}_{[\tilde{t}'_0] [t_k]} = -
\tilde{\gamma}r^{\frac{1}{2}}\delta^{k1}$ for
$\tilde{\gamma}^2=-1$.} In their turn, the original
$\mathcal{A}_b$ and modified $\mathcal{A}_{b{}mod}$ nonlinear
massive (massless, $m=0$) integer-spin algebras for bosonic HS
fields, i.e., tensors in (\ref{Young k2}), for $s_i= n_i$, contain
only the respective bosonic elements $o_{I_b}$, with
$\tilde{o}_{I_b}$ being Lorentz scalar having no $\gamma$-matrices
in the definition of $D_{\mu}$ as compared to the
$\bigl(2^{[\frac{d}{2}]}\times 2^{[\frac{d}{2}]}\bigr)$-matrix
structure of $o_{I}$, $\tilde{o}_{I}$ (with $[x]$ being fractional
part  of a number $x \in \mathbb{R}$) for the superalgebras
$\mathcal{A}$, $\mathcal{A}_{mod}$, and obey, in the case of
$\tilde{o}_{I_b}$, the same algebraic relations as those given by
Table~\ref{til-table} without the fermionic operators $t_0, t_i,
t_i^+$. Only some of the nonlinear relations (\ref{lilj}),
(\ref{l'il'j+}) are changed: first, $t^i$ must be removed from
(\ref{lilj}), second, Eqs.~(\ref{l'il'j+}), along with the new
definition of $l_{ob} = \bigl(D^2 -
r\textstyle\frac{d(d-6)}{4}\bigr)\in\{\tilde{o}_{Ib}\},
\tilde{l}_{ob}\in\{{o}_{Ib}\}$, acquire the form
\begin{eqnarray}
 {} {X}^{ij}_b
 \hspace{-0.3em} & \hspace{-0.3em}= \hspace{-0.3em}
 &\hspace{-0.3em} \Bigl\{{{l}}_{0b}
+ r\Bigl(\textstyle  (g_0^i -1 - \delta^{i1})
{g_0^i}-(1+\delta^{i2}){g_0^2} +t^+t\Bigl)\Bigl\}\delta^{ij} -
r\Bigl\{ 4\sum_{k}
l^{jk+}l^{ik}\nonumber \\
{}& &  + (g_0^1+g_0^2 -{2})t\delta^{j1}\delta^{i2} +
t^+(g_0^1+g_0^2 -{2})\delta^{j2}\delta^{i1} \Bigl\}
\label{lilj+b},\\
\tilde{l}_{ob} \hspace{-0.3em} & \hspace{-0.3em}= \hspace{-0.3em}
 &\hspace{-0.3em}  {l}_{ob} + \tilde{m}^2_b + r \bigl((g_0^1-2\beta-2)g_0^1 -
 g_0^2 \bigr), \ \ \tilde{m}^2_b = {m}^2 + r\beta(\beta+1),\label{l0b}
\end{eqnarray}
being a consequence of the AdS-group irrep equations for the tensor
$\Phi_{(\mu)_{s_1},\ (\nu)_{s_2}}$, see \cite{Metsaev},
\begin{eqnarray}
\label{Eq-0b} &&\bigl[\nabla^2 +r[(s_1-\beta-1+ d)(s_1-\beta) -
s_1- s_2]+m^2 \bigr]\Phi_{(\mu)_{s_1},\ (\nu)_{s_2}} =0,\\
&&\bigl(\ g^{\mu_1\mu_2} ,\ g^{\nu_1\nu_2},\ g^{\mu_1\nu_1}
\bigr)\Phi_{(\mu)_{s_1},\ (\nu)_{s_2}} =
\Phi_{\{(\mu)_{s_1},\nu_1\}\nu_2...\nu_{s_2}}=0, \label{Eq-1b}
\end{eqnarray}
realized in $\mathcal{H}$, with a standard scalar product
$\langle\ |\ \rangle$, as constraints: $(\tilde{l}_{ob}, l_{ij},
t)|\Phi\rangle=0$.

\vspace{2ex} {\center\subsubsection{ADDITIVE CONVERSION FOR
NONLINEAR SUPERALGEBRAS AND VERMA MODULE
CONSTRUCTION}}\label{addconvVermamod} \vspace{1ex}

To convert additively non-linear superalgebras with a subset
of 2nd class constraints, we need the following easily verified

\emph{\underline{Proposition}:} If a set of operators
$\{\tilde{o}_I\}, \{\tilde{o}_I\}: \mathcal{H} \to \mathcal{H}$ is
subject to $n$-th order polynomial supercommutator relations (with
the Grassmann parities $\varepsilon_I$=$\varepsilon(o_I)$=$0,1$)
\begin{eqnarray}
[\,\tilde{o}_I,\tilde{o}_J\} &=& f_{IJ}^{K_1}\tilde{o}_{K_1} +
\sum_{m=2}^nf_{IJ}^{K_1\cdots K_m}\prod_{l=1}^{m}\tilde{o}_{K_l}
,\ f_{IJ}^{K_1\cdots K_m} = -(-1)^{\varepsilon_I\varepsilon_J}
f_{JI}^{K_1\cdots K_m}, \label{inal}
\end{eqnarray}
then due to the requirement of the composition law for a direct sum,
$$ \tilde{o}_J \longrightarrow
\tilde{O}_J = \tilde{o}_J + \tilde{o}'_J:  \{\tilde{o}'_I\}:
\mathcal{H}' \to \mathcal{H}',\quad  [\tilde{o}_I, \tilde{o}'_J\}
=0,\quad \mathcal{H} \bigcap \mathcal{H}' = \emptyset,
$$
such that the set of enlarged operators $\{\tilde{O}_I\}$ must
obey involution relations, {$ [\tilde{O}_I,\tilde{O}_J\}$ = $
\tilde{F}_{IJ}^K(\tilde{o}',\tilde{O}) \tilde{O}_K $}, the sets
$\{\tilde{o}'_J\}$, $\{\tilde{O}_J\}$ form nonlinear superalgebras
$\mathcal{A}'$, $\mathcal{A}_{c}$, given in $\mathcal{H}'$ and
$\mathcal{H}\otimes \mathcal{H}'$ with the corresponding explicit
multiplication laws
\begin{eqnarray} \label{addalg}
\hspace{-1,5ex}&&  [\,o_I',o_J'\} = f_{IJ}^{K_1}{o}'_{K_1}+
\sum_{l=2}^n(-1)^{
  l-1 +\varepsilon_{K_{(l)}}}f_{IJ}^{K_{l}\cdots
K_1}\prod_{s=1}^{l}{o}'_{K_s},\
\\
\label{conv-alg}
 \hspace{-1,5ex} && [\,\tilde{O}_I,\tilde{O}_J\} \hspace{-0,3ex}=\hspace{-0,3ex}
  \Bigl(f_{IJ}^{K} + \sum_{l=2}^{n}
  F^{(l){}K}_{IJ}({o}',\tilde{O})\Bigr)\tilde{O}_{K},\ \varepsilon_{K_{(n)}} =
\sum_{s=1}^{n-1}\varepsilon_{K_s}\Bigl(\sum_{l=s+1}^n\varepsilon_{K_l}\Bigr),
\\
 \hspace{-1,5ex} && F^{(l){}K_l}_{IJ}
  \hspace{-0,3ex}=\hspace{-0,3ex}
f_{IJ}^{K_1\cdots K_{l}}\prod_{m=1}^{l-1}\tilde{O}_{K_m}
+\sum_{s=1}^{l-1}(-1)^{s+ \varepsilon_{K_{(s)}}}
f_{ij}^{\widehat{K_s\cdots K_1}\widehat{K_{s+1}\cdots K_{l}}}
\prod_{p=1}^{s}{o}'_{K_{p}}
\prod_{m=s+1}^{l-1}\tilde{O}_{K_{m}},\nonumber \\
 \hspace{-1,5ex}&& f_{ij}^{\widehat{K_s\cdots K_1}\widehat{K_{s+1}\cdots K_{l}}} =
f_{ij}^{{K_s\cdots K_1}{K_{s+1}\cdots K_{l}}} + f_{ij}^{K_s\cdots
{K_{s+1}K_1}{K_{s+2}\cdots
K_{l}}}(-1)^{\varepsilon_{K_{s+1}}\varepsilon_{K_1}} +\cdots +
\nonumber \\
 \hspace{-1,5ex}{}&{}&  f_{ij}^{{K_{s+1}K_s\cdots K_1}{K_{s+2}\cdots
K_{l}}}(-1)^{\varepsilon_{K_{s+1}}\sum_{l=1}^{s}\varepsilon_{K_l}}
+\Bigl(f_{ij}^{K_{s+1}K_s\cdots {K_{s+2}K_1}{K_{s+3}\cdots
K_{l}}}(-1)^{\varepsilon_{K_{s+2}}\varepsilon_{K_1}} +\nonumber \\
 \hspace{-1,5ex}{}&{}&\cdots +
f_{ij}^{{K_{s+1}K_{s+2}K_s\cdots K_1}{K_{s+3}\cdots
K_l}}(-1)^{\varepsilon_{K_{s+2}}\sum_{l=1}^{s}\varepsilon_{K_l}}\Bigr)(-1)^{\varepsilon_{K_{s+1}}\sum_{l=1}^{s}\varepsilon_{K_l}
}+ \cdots
+\nonumber\\
 \hspace{-1,5ex}{}&{}&
(-1)^{\sum_{m=s+1}^{l}\varepsilon_{K_{m}}\sum_{l=1}^{s}\varepsilon_{K_l}
}f_{ij}^{{K_{s+1}\cdots K_l}K_s\cdots K_1},\label{sumcoeff}
\end{eqnarray}
where the sum (\ref{sumcoeff}) contains  $\frac{l!}{s!(l-s)!}$
terms with all the possible ways of ordering the indices
$(K_{s+1},..., K_{l})$ among the indices $(K_{s},..., K_{1})$ in
$f_{ij}^{{K_s\cdots K_1}{K_{s+1}\cdots K_{l}}}$.

As a consequence, for $n=2$ in Proposition, as well as for the
algebraic relations given by Table~\ref{til-table} for
$(\mathcal{A})\mathcal{A}_b$, we can obtain relations (for the
first time deduced in \cite{adsfermBKR} for quadratic
superalgebras) for the (super)algebras
$(\mathcal{A}')\mathcal{A}'_b$ of the additional $o'_I$ and for
the (super)algebras $(\mathcal{A}_c)\mathcal{A}_{b{}c}$ of the
converted operators $\tilde{O}_I$. These relations remain the same
for the linear (Lie) part of the superalgebras, with the only
respective change $\tilde{o}_I \rightarrow ({o}'_I, \tilde{O}_I)$,
whereas the quadratic ones (\ref{t0li})--(\ref{lilj+b}) take  the
form (with a preservation of Table~\ref{til-table}, except for the
replacement $({\mathcal{K}}^{bi}_1, {\mathcal{K}}^{bi+}_1, {M}^{
i}, {M}^{ i+}) \rightarrow-({\mathcal{K}}^{\prime bi}_1,
{\mathcal{K}}^{\prime bi+}_1, {M}^{\prime i}, {M}^{\prime i+})$,
for $\mathcal{A}'$, $(r{\mathcal{K}}^{bi}_1,r
{\mathcal{K}}^{bi+}_1, {M}^{ i}, {M}^{ i+})\rightarrow (-V^{i+}_W,
-V^i_W, \hat{M}^{ i}_W, \hat{M}^{ i+}_W)$ for $\mathcal{A}_{c}$
\begin{eqnarray}{}
{M}^{\prime i} &=& - r\Bigl(2\textstyle\sum_{k}t^{
  \prime k+}l^{\prime ki}
+g_0^{\prime i}t^{\prime i} -{\textstyle\frac{1}{2}}t^{\prime i}
-t't^{\prime 1}\delta^{i2}- t^{\prime +}t^{\prime 2}\delta^{i1}
\Bigr)   , \label{t'0l'i}
\\
{W}^{\prime ij} & = &{W}^{ \prime ij}_b -
\textstyle\frac{r}{4}t^{\prime [j}t^{\prime i]}  =
-2r\varepsilon^{ij}\left[(g_0^{\prime 2}-g_0^{\prime 1})l^{\prime
12} - t' l^{\prime 11} + t^{\prime +} l^{\prime 22}\right] -
\textstyle\frac{r}{4}t^{\prime[j}t^{\prime i]}   ,
 \label{l'il'j}\\
{\mathcal{K}}^{\prime bi}_1& = &\Bigl(4\textstyle\sum_{k}l^{\prime
ik+}l^{\prime k} +l^{\prime i+}(2g_0^{\prime i}-1) -2l^{\prime
2+}t' \delta^{i1} - 2l^{\prime 1+}t^{\prime +}
\delta^{i2}\Bigr)\label{l'0l'i+} ,\\
 {} {X}^{\prime ij}
 \hspace{-0.5em} & \hspace{-0.5em}= \hspace{-0.5em}
 &\hspace{-0.5em} \Bigl\{\hspace{-0.1em}{l}'_0
- r\Bigl(\hspace{-0.2em}\textstyle\sum_{k}K^{\prime1k}_0 +
K^{\prime0i}_0 + \frac{1}{2}K^{\prime 1i}_0 +
\mathcal{K}^{\prime12}_0
\hspace{-0.2em}\Bigr)\hspace{-0.2em}\Bigr\}\delta^{ij} +
r\Bigl\{\hspace{-0.2em} \Bigl[\hspace{-0.2em}4\sum_{k} l^{\prime
1k+}l^{\prime k2}
\hspace{-0.1em}-\hspace{-0.1em}\textstyle\frac{1}{2}t^{\prime
1+}t^{\prime 2}\nonumber
\\
%
%
%
 \hspace{-0.5em} & \hspace{-0.5em} \hspace{-0.5em}
 &\hspace{-0.5em}  + \textstyle(\sum_{k}g_0^{\prime k} -\textstyle\frac{3}{2}
)t'\hspace{-0.15em}\Bigr]\hspace{-0.15em}\delta^{j1}\delta^{i2}
\hspace{-0.15em}+\hspace{-0.15em}\Bigl[\hspace{-0.2em}4\sum_{k}
l^{\prime k2+}l^{\prime 1k}-\textstyle\frac{1}{2}t^{\prime
2+}t^{\prime 1}+ t^{\prime +}(\sum_{k}g_0^{\prime k} -
\textstyle\frac{3}{2})\hspace{-0.1em}\Bigr]
\hspace{-0.2em}\Bigr\}\hspace{-0.1em}
\label{fl'il'j+},\\
 {} {X}^{\prime ij}_b \hspace{-0.5em} & \hspace{-0.5em}=
\hspace{-0.5em}
 &\hspace{-0.5em}  \textstyle\Bigl\{{l}'_0 - r\Bigl(K^{\prime 0i}_0 +
\mathcal{K}^{\prime 12}_0\Bigr)\Bigl\}\delta^{ij} + r\Bigl\{
\Bigl[4\sum_{k} l^{1k+}l^{k2}
+ (\sum_{k}g_0^k - {2})t\Bigr]\delta^{j1}\delta^{i2} \nonumber \\
{}& & + \textstyle\Bigl[4\sum_{k} l^{k2+}l^{1k}+ t^+(\sum_{k}g_0^k
 - {2})\Bigr]\delta^{j2}\delta^{i1} \Bigl\}.
\label{li'lj'+ib}
\end{eqnarray}
In their turn, the only modified relations, for instance, in the
converted algebra $\mathcal{A}_{bc}$, have the form (with
the choice of Weyl's ordering of $\tilde{O}_{I_b}$ in the r.h.s.
of the commutators), which implies, as in \cite{adsfermBKR},
an exact expression for the BRST operator,
\begin{eqnarray}
{{V}}^{i+}_{bW}\hspace{-0.5em}& \hspace{-0.3em}=
\hspace{-0.3em}&\hspace{-0.5em} - r\Bigl(2(L^{ ii+}-2l^{\prime
ii+})L^{ i}+ 2(L^{i}-2l^{\prime i})L^{ ii+} + (L^{i+}-2l^{\prime i
+})G^{i}_0 + (G^{i}_0-2g^{\prime i}_0) L^{ i +}\nonumber \\
\hspace{-0.5em}&\hspace{-0.5em}&\hspace{-0.5em} +2
\bigl[\bigl((L^{12 +}-2l^{\prime12 +})L^{ \{1}  +(L^{
\{1}-2l^{\prime \{1})L^{12 +}\bigr)\delta^{2\}i }-
\textstyle\frac{1}{2}\delta^{1i}\bigl((L^{2+}-2l^{\prime 2+})T
\nonumber \\
\hspace{-0.5em}&\hspace{-0.5em}&\hspace{-0.5em} +(T- 2t')L^{
2+}\bigr) - \delta^{2i}\bigl((L^{1+}-2l^{\prime 1+})T^{ +} +
(T^+-2t^{\prime +})L^{ 1+}\bigr)\bigr]\Bigr) \label{L0Li+bW} ,
\\
{} {{W}}^{ij}_{bW} \hspace{-0.5em}&\hspace{-0.3em} =
\hspace{-0.3em}&\hspace{-0.5em}
r\varepsilon^{ij}\Bigl\{\textstyle\sum_k(-1)^k(G_0^k -
2g_0^{\prime k}) L^{12}+(L^{12}-2l^{\prime 12})\sum_k(-1)^kG_0^k -
[ (T- 2t') L^{11} \nonumber \\
{}\hspace{-0.5em}&\hspace{-0.5em} &   \hspace{-0.5em}  + (L^{11} -
2l^{\prime 11})T]  + (T^+-2t^{\prime +}) L^{22} + (L^{22}
-2l^{\prime 22})T^+ \Bigr\} ,
 \label{LiLjbW}  \\
{} {\hat{X}}^{ij}_{bW} \hspace{-0.5em}&\hspace{-0.3em} =
\hspace{-0.3em}& \hspace{-0.5em} \Bigl\{L_0 + r\bigl((G_0^i -
2g^{\prime i}_0)G_0^i +\textstyle \frac{1}{2}\{T^+,T\} -
(t^{\prime+}T+ t'T^+) \bigr)\Bigl\}\delta^{ij}-r\Bigl\{ 2\sum_k
[(L^{jk+}
 \nonumber\\
\hspace{-0.5em}&\hspace{-0.5em}\hspace{-0.5em}&\hspace{-0.5em}
\textstyle -2l^{\prime jk+})L^{ik} +(L^{ik}- 2l^{\prime
ik})L^{jk+}] + \frac{1}{2}\bigl[\sum_{k}(G_0^k -2g_0^{\prime k})
T +(T - 2t')\hspace{-0.4em} \times\nonumber \\
{}\hspace{-0.5em}&\hspace{-0.5em} &\hspace{-0.5em}
\textstyle\times\sum_{k}G_0^k\bigr]\delta^{j1}\delta^{i2} +
\frac{1}{2}\bigl[(T^+ - 2t^{\prime +})\sum_{k}G_0^k +
\sum_{k}(G_0^k -2g_0^{\prime k})T^+\bigr]\delta^{j2}\delta^{i1}
\Bigr\}
 \label{LiLj+bW} .
\end{eqnarray}
In Eqs.~(\ref{fl'il'j+}), (\ref{li'lj'+ib}), we have presented
the quantities $K^{0i}_0$, $K^{0i}_0 = ({g_0^i}^2-2g_0^i -
4l^+_{ii})$, being Casimir operators for the bosonic
subalgebras $so(2,1)$ generated by $l_{ii}, l_{ii}^+, g_0^i$
for each $i=1,2$. The operators $K^{1i}_0$, $K^{1i}_0 = (g_0^i +
t^{i+}t^i)$ extend $K^{0i}_0$ up to the Casimir operators
$\mathcal{K}_0^i$, $\mathcal{K}_0^i = ({K}_0^{0i}+{K}_0^{1i})$, of
the Lie subsuperalgebras in $\mathcal{A}_c$ generated by
$ (t^i,t^{i+}, l_{ii}, l_{ii}^+, g_0^i)$ for each $i=1,2$, and the
quantity $\mathcal{K}^{12}_0$ extend $\sum_i{K}_0^{0i}$,
$\sum_i\mathcal{K}_0^i$ up to the respective Casimir operators
$\mathcal{K}_0$, $\mathcal{K}_0^b$,
\begin{eqnarray}\label{K_0}
    \mathcal{K}_0 & = & \mathcal{K}^b_0 + \sum_{i}K_0^{1{}i} =
    \sum_{i}\bigl(K_0^{0{}i} + K_0^{1{}i}\bigr) +
    2\mathcal{K}_0^{12},\
    \mathcal{K}_0^{12} =
    t^{\prime +}t' - g_0^{\prime 2} - 4l^{\prime 12 +}
      l^{\prime 12},
\end{eqnarray}
of the maximal (in $\mathcal{A}$) Lie superalgebra
$\mathcal{A}^{Lie}$ generated by $(t^i, t^{i+}, l_{ik}, l_{ik}^+,
g_0^i, t, t^+), i, k=1,2$, and of its $so(3,2)$ subalgebra.

These operators appear to be crucial to realize the operators
of the (super)algebra $(\mathcal{A}')\mathcal{A}'_b$ in terms
of the creation and annihilation operators of a new Fock space
$(\mathcal{H}')\mathcal{H}'_b$, whose number of pairs is equal to
that of the converted 2nd-class constraints $o_\mathrm{a}$, which
allows one to obtain the correct number of physical degrees
of freedom describing the basic spin-tensor (\ref{Young k2}) in the
final LF, after an application of the BFV--BRST procedure to
the resulting first-class constraints $\{\tilde{O}_\alpha\}
\subset\{\tilde{O}_I\}$.

Among the two variants of an additive conversion for the non-linear
superalgebras \cite{0809.4815SMPTP} of $\{o_I\}$ into the
1st-class system $\{O_\alpha\}$ [first, for the total set
of $\{o_\mathbf{a}\}$, resulting in an unconstrained LF,
second, for the differential and partly algebraic constraints
$l_i, l_i^+,t, t^+$, restricting the (super)algebra $\mathcal{A}$
to the surface $\{o^r_\mathbf{a}\}\equiv \{o_\mathbf{a}\}\setminus \{l_i,
l_i^+, t, t^+\}$ at all the stages of the construction, resulting
in an LF with off-shell $\gamma$-traceless (only for the fermionic HS
field $\Phi_{(\mu)_{n_1},(\nu)_{n_2}}$) and (only) traceless
conditions for the fields and gauge parameters], we consider in
detail the former case.
To find $o'_I$ explicitly, we need, first, to construct
an auxiliary representation, known as the Verma module
\cite{Dixmier}, on the basis of a Cartan-like decomposition,
extended from the one for $\mathcal{A}^{Lie}$,
\begin{equation}\label{Cartandecomp}
    \mathcal{A}' =  \{\{t^{\prime+}_i, l^{\prime ij+},
t^{\prime+}; l^{\prime i+}\} \oplus \{g_0^{\prime i}; t_0', l_0'\}
\oplus \{t'_i,l^{\prime ij}, t';l^{\prime i}\} \equiv
\mathcal{E}^-\oplus H \oplus\mathcal{E}^+,
\end{equation}
and then to realize the above Verma module as an operator-valued
formal power series $\sum_{n\geq 0}\sqrt{r}^n$
$\mathcal{P}_n[(a,a^+)_\mathbf{a}]$ in a new Fock space
$\mathcal{H}'$ generated by $(a,a^+)_\mathbf{a} = f_i,f^{+}_i
b_i$, $b^{+}_i, b_{ij}$, $b_{ij}^+, b, b^+$ (for a constrained LF,
$\{o^r_\mathbf{a}\} \leftrightarrow (a,a^+)^r_\mathbf{a} = \{ b_i,
b^{+}_i, b, b^+$\}).

A solution of these problems is more involved than the analysis
made for a non-linear $\mathcal{A}'$, see \cite{adsfermBKR}, and
for a Lie superalgebra $\mathcal{A}^{\prime Lie}$, see \cite{flatfermmix},
due to a nontrivial entanglement of the triplet of non-commuting
negative root vectors $\bigl({l_1^{\prime +}}t^{\prime +}{l_2^{\prime
+}}\bigr)$ and their ordered products $\bigl(({l_1^{\prime
+}})^{n_1}(t^{\prime +})^n({l_2^{\prime +}})^{n_2}\bigr)$,
$n_i,n\in \mathbb{N}_0$ (composing the non-commuting part of an
arbitrary vector of the Verma module $V_{\mathcal{A}'}$), as
follows from Table~\ref{til-table}. This task should be effectively
solved iteratively, e.g., for an action of the operator $t'$ on
$({l^{\prime 2+}})^{n_2}$,
\begin{equation}\label{t'action}
t'\bigl({l^{\prime 2+}}\bigr)^{n_2} \rightarrow \sum_{m_j=0}
\bigl({l^{\prime 2+}}\bigr)^{n_2-1-2m_j}l^{\prime 1+} + ...
\rightarrow t'\sum_{m'=0}\sum_{m=0} \bigl({l^{\prime
2+}}\bigr)^{n_2-2-2m-2m'} + ...\,,\nonumber
\end{equation}
where the remaining summand does not contain any incorrectly
ordered terms, thus, extending the known results of Verma module
construction \cite{0206027} and its Fock space realization in
$\mathcal{H}'$. First, note that there is no nontrivial
entanglement of the above triplet of negative root vectors, due to
a restriction of Table~\ref{til-table} to the surface determined
by the non-converted second-class constraints $l^{ij}, l^{ij+},
t^i,t^{i+}$, and, therefore, the finding of an operator
realization of the restricted $\mathcal{A}'_r$ follows the known
way \cite{adsfermBKR}. Second, within our conversion procedure the
enlarged central charge $\tilde{M} = \tilde{m} + \tilde{m}'$
vanishes, whereas explicit expressions for $o'_I$ in terms of
$(a,a^+)_\mathbf{a}$ and the new constants $m_0, h^i$, $(l'_0,
g^{\prime i}_0) = (m_0^2, h^i) + ...$ [they are to be determined
later from the condition of reproducing the correct form of
Eqs.~(\ref{t't0})] are found by partially following
\cite{adsfermBKR, 0206027}.

\vspace{2ex} {\center\subsubsection{BFV-BRST OPERATOR FOR
CONVERTED
 (SUPER)ALGEBRA}}\label{convBFVBRST}
\vspace{1ex}

  To construct a BRST operator for a non-linear
non-gauge (super)algebra $(\mathcal{A}_c)\mathcal{A}_{b{}c}$, we
shall use an operator version of finding a BRST operator,
described in Ref. \cite{BFV2}, and classically in Ref.
\cite{Henneaux}. Due to the quadratic algebraic relations
(\ref{L0Li+bW})--(\ref{LiLj+bW}) and their Her\-mi\-ti\-an
conjugates, we must check a nontrivial existence of new structure
relations and new structure functions of 3rd order
\cite{Henneaux}, implied by a resolution of the Jacobi
identities $(-1)^{\varepsilon_I\varepsilon_K}[[\tilde{O}_I,
\tilde{O}_J\},\tilde{O}_K\} + cycl. perm. (I,J,K)=0$, for
$(\mathcal{A}_c)\mathcal{A}_{b{}c}$, $n=2$ in (\ref{conv-alg}),
(\ref{sumcoeff}),
\begin{eqnarray}\label{Jid}
   &&  (-1)^{\varepsilon_I\varepsilon_K}\Bigl(\bigl(f^{M}_{IJ}+
   F^{(2){}M}_{IJ}\bigr)\bigl(
   f^{P}_{MK} + F^{(2){}P}_{MK} \bigr) + (-1)^{\varepsilon_P\varepsilon_K}[F^{(2){}P}_{IJ},
   \tilde{O}_K\}\Bigr)
     \nonumber\\
   &&+ cycl.perm.(I,J,K)-
   \textstyle\frac{1}{2} F_{IJK}^{RS} (f^{P}_{RS}+ F^{(2){}P}_{RS})  =
   F_{IJK}^{RP}\tilde{O}_R ,   \nonumber\\
   &&
            F^{(2){}K}_{IJ}({o}',\tilde{O})=
      -\bigl(f_{IJ}^{MK}+(-1)^{\varepsilon_K\varepsilon_M}
      f_{IJ}^{KM}\bigr)o_M' + f_{IJ}^{MK}\tilde{O}_M,
\end{eqnarray}
with the 3rd order structure functions
$F_{IJK}^{RS}({o}',\tilde{O})$ satisfying the properties of
generalized antisymmetry with respect to a permutation of any two
of the lower indices $(I,J,K)$ and the upper indices
$R,S$.\footnote{Given by Eqs.~(\ref{Jid}), the resolution of
Jacobi identities for a nonlinear superalgebra is more general
than the one presented in Ref.~\cite{0701243} for a classical
(super)algebra, because we do not examine a more restrictive
vanishing of all the coefficients at the 1st, 2nd and 3rd degrees
in $\tilde{O}_I$} If the 4th-, 5th- and 6th- order structure
functions $F_{IJKL}^{PRS}({o}',\tilde{O})$,
$F_{IJKLM}^{PRST}({o}',\tilde{O})$,
$F_{IJKLMN}^{PRSTU}({o}',\tilde{O})$ are zero, the BRST operator
$Q'$ has the form of the one for a formal 2nd-rank ``gauge''
theory \cite{Henneaux}, i.e., it has an exact form for the
$(\mathcal{C} \mathcal{P})$-ordering of the ghost coordinates
$\mathcal{C}^I$, bosonic, $q_0, q_i,q_i^+$, and fermionic,
$\eta_0$, $\eta_i^+$, $\eta_i$, $\eta_{ij}^+$, $\eta_{ij}$,
$\eta$, $\eta^+$, $\eta^i_G$, and their conjugated momenta
operators $\mathcal{P}_I$: $p_0$, $p_i^+$, $p_i$, ${\cal{}P}_0$,
${\cal{}P}_i$, ${\cal{}P}_i^+$, ${\cal{}P}_{ij}$,
${\cal{}P}_{ij}^+$, $\mathcal{P}$, $\mathcal{P}^+$,
${\cal{}P}^i_G$, see \cite{flatfermmix}, with the Grassmann
parities opposite to those of $\tilde{O}_I$ and the values of
ghost number $gh(\mathcal{C}^I) = -gh(\mathcal{P}_I)=1$ with only non-vanishing graded commutators $[\mathcal{C}^I,\,\mathcal{P}_J\}=\delta^I_J$,
\begin{eqnarray}\label{genQ'}
    {Q}'  = \mathcal{C}^I\bigl[{\tilde{O}}_I  + \textstyle\frac{1}{2}
    \mathcal{C}^{J}(f^{P}_{JI}+
   F^{(2){}P}_{JI})\mathcal{P}_{P}
    (-1)^{\varepsilon_{I}+\varepsilon_{P}}+\frac{1}{12}
    \mathcal{C}^{J}\mathcal{C}^{K}
    F^{RP}_{KJI}\mathcal{P}_{R}\mathcal{P}_{P}(-1)^{\varepsilon_{I}
    \varepsilon_{K}+\varepsilon_{J}+\varepsilon_{R}}\bigr]\hspace{-0.1em}.
\end{eqnarray}
The requirement of $(\mathcal{C} \mathcal{P})$-ordering  for the
equations  following from the nilpotency condition for ${Q'}$ in
the $n$th- order  in  $\mathcal{C}^I$, $n=3,4,5,6$, leads, for
instance, for $n=3$ to the necessity or the separate
fulfillment the relations which do not appear in the classical
case \cite{Henneaux}:
\begin{eqnarray}\label{Jiddifcons}
     (-1)^{\varepsilon_I\varepsilon_K}\bigl[F^{(2){}M}_{IJ},
   F^{(2){}P}_{MK} \bigr\} + cycl.perm.(I,J,K)-
   \textstyle\frac{1}{2} \bigl[F_{IJK}^{RS}, F^{(2){}P}_{RS}\bigr\}  =
   \bigl[F_{IJK}^{RP},\tilde{O}_R\bigl\} ,
\end{eqnarray}
or  a corresponding extension  of the 3rd structural relations
(\ref{Jid}) by these terms.

 In the case of a bosonic algebra $\mathcal{A}_{bc}$, there
are 3 types of nontrivial Jacobi identities for 6 triplets $(L_1,
L_2, L_0)$, $(L_1^+, L_2^+, L_0)$, $(L_i, L_j^+, L_0)$, with the
existence of 3rd-order structure functions. For instance, one of
the solutions for $(L_i, L_j^+, L_0)$ after a reduction of
$L_{11}^+$ has the form
\begin{eqnarray}
&&2\Bigl\{\delta^{i2}\delta^{j1}\Bigl[(L^{22}-2l^{\prime
22})(T^{+}-2t^{\prime +}) + (G_0^i-2g_0^{\prime
i})(L^{12}-2l^{\prime 12})  \nonumber\\
 &&  + r^{-1}(\hat{W}^{ij}_{bW}-2{W}^{\prime
ij}_{b})- (T-2t^{\prime })(L^{11}-2l^{\prime 11})
-(G_0^j-2g_0^{\prime j})(L^{12}-2l^{\prime 12})\Bigr] \nonumber\\
 && -
\varepsilon^{\{1j}\delta^{2\}i}\Bigl[ (L^{12}-2l^{\prime 12})
 (T^{+}-2t^{\prime +})-(T^{+}-2t^{\prime +})
 (L^{12}-2l^{\prime 12}) \Bigr]\Bigr\}\nonumber\\
 &=&\delta^{i2}\delta^{j1}\Bigl(\{L_{11},T\}-\{L_{22},T^+\} -
 \{L_{12},G_0^2-G_0^1\} - 4L^{12}\Bigr)
 +2\varepsilon^{\{1j}\delta^{2\}i}L^{11}.
\end{eqnarray}
As a result, in view of the absence of higher-order structure
functions, a nilpotent BRST operator $Q'$ (\ref{genQ'}) for
$\mathcal{A}_{bc}$ has an exact form of the maximal 3rd degree
in the powers of ghosts $\mathcal{C}^I$:
\begin{eqnarray}
\label{Q'b} \hspace{-3em}  \hspace{-0.4em}
&Q'\hspace{-0.4em}&\hspace{-0.4em} = \textstyle
\frac{1}{2}\eta_0L_0+\eta_i^+L^i +\eta_{lm}^+L^{lm}  + \eta^+T  +
\frac{1}{2}\eta^i_{{G}}{G}_i
+\frac{\imath}{2}\eta_i^+\eta^i{\cal{}P}_0 + \frac{\imath}{2}
\eta_{ii}^+\eta^{ii}{\cal{}P}^i_{{G}}+\frac{\imath}{2}\eta_i^+\eta^i{\cal{}P}_0\nonumber
\\\hspace{-0.4em}&&
{}\hspace{-0.4em}   + \textstyle\frac{\imath}{2}
\eta_{ii}^+\eta^{ii}{\cal{}P}^i_{{G}} + (
\eta^i_{{G}}\eta_i^++\eta_{ii}^+\eta^i){\cal{}P}^i+2\eta^i_{{G}}
\eta_{ii}^+{\cal{}P}_{ii} - \eta_{12}(\eta^+\mathcal{P}_{11}^+ +
\eta\mathcal{P}_{22}^+)\nonumber
\\
\hspace{-0.4em}&&{}\hspace{-0.4em}  -2
\left[\textstyle\frac{1}{2}\sum_k\eta^k_{{G}}\eta_{12}
-\eta^+\eta_{22} -\eta\eta_{11}\right]\mathcal{P}_{12}^+ +
\textstyle\frac{\imath}{2} \eta\eta^+\sum_k(-1)^k{\cal{}P}^k_{{G}}
+ \textstyle
\frac{\imath}{8}\eta_{12}^+\eta_{12}\sum_k{\cal{}P}^k_{{G}}
\nonumber\\
\hspace{-0.4em} && \hspace{-0.4em} +
\bigl[\textstyle\frac{1}{2}\eta_{12}^+\eta_{11}+
\textstyle\frac{1}{2}\eta_{22}^+\eta_{12}  +
\sum_k(-1)^k\eta^k_{{G}}\eta^+\bigr]\mathcal{P} +
\bigl[\textstyle\frac{1}{2}\eta_{12}^+\eta_2  - \eta
\eta_2^+\bigr]\mathcal{P}_1  +
\bigl[\textstyle\frac{1}{2}\eta_{12}^+\eta_1 - \eta^+
\eta_1^+\bigr]\mathcal{P}_2   \nonumber
\end{eqnarray}
\vspace{-5ex}
\begin{eqnarray}
\label{Q'nlb2}\hspace{-0.4em}&\hspace{-0.4em}+
\hspace{-0.4em}&\hspace{-0.4em} r\Bigl\{
 \eta_0 \eta^+_i\bigl(
 2(L^{ ii}-2l^{\prime
ii})\mathcal{P}_i^+ + 2(L^{ i+} - 2l^{\prime i+})\mathcal{P}_{ii}
 - \imath (L^i - 2l^{\prime i})\mathcal{P}_G^{i}
 + (G^{i}_0 -2 g^{\prime i}_0) \mathcal{P}_{ i }\nonumber \\
{}&& +2 \bigl[\bigl((L^{12 }-2l^{\prime12 })\mathcal{P}^{\{1+} +
(L^{ \{1+}-2l^{\prime \{1+})\mathcal{P}_{12 }\bigr)\delta^{2\}i }
-\textstyle\frac{1}{2}\delta^{1i}\Bigl((L^{2}-2l^{\prime2})\mathcal{P}^+
 \nonumber \\
&&  + (T^+ - 2 t^{\prime +})\mathcal{P}_{2}\Bigr) - \textstyle\frac{1}{2}\delta^{2i}
\Bigl((L^{ 1}-2l^{\prime
1})\mathcal{P} + (T-2t^{\prime
})\mathcal{P}_{1}\bigr)\bigr]\bigr)\nonumber
\\
&& - \textstyle\frac{1}{2}\eta^+_i \eta^+_{j}
\varepsilon^{ij}\bigl\{\sum_k(-1)^k(G_0^k - 2g_0^{\prime
k})\mathcal{P}_{12}-\imath(L^{12}-2l^{\prime 12})\sum_k(-1)^k
\mathcal{P}_G^k \nonumber \\
{}& &   -[ (T- 2t') \mathcal{P}_{11} + (L^{11} - 2l^{\prime
11})\mathcal{P}]  + (T^+-2t^{\prime +}) \mathcal{P}_{22} + (L^{22}
-2l^{\prime
22})\mathcal{P}^+ \bigr\}\nonumber \\
&  & +\textstyle2\eta^+_i \eta_{j}\bigl\{\sum_{k}
(L^{jk+}-2l^{\prime jk+})\mathcal{P}^{ik}+ \bigl[ \textstyle
\frac{\imath}{4} (G_0^i-2g_0^{\prime
i})\mathcal{P}_G^i -\frac{1}{8}(T^+-2t^{\prime +})\mathcal{P}   \nonumber \\
{}& & \textstyle-\frac{1}{8}(T-2 t')\mathcal{P}^+
\bigr]\delta^{ij} + \frac{1}{4}\bigl[\sum_{k}(G_0^k -2g_0^{\prime
k}) \mathcal{P} -\imath (T -
2t')\sum_{k}\mathcal{P}_G^k\bigr]\delta^{j1}\delta^{i2} \bigr\}
\Bigr\}\nonumber \\
 &+& r^2\eta_0\eta_i\eta_j\varepsilon^{ij}\Bigl\{
\textstyle\frac{1}{2}( \sum_kG^k_0[\mathcal{P} \mathcal{P}^{22+}
 - \mathcal{P}^+
\mathcal{P}^{11+}] -
  \frac{i}{2}({L}^{11+}\mathcal{P}^+-{L}^{22+}\mathcal{P})
  \sum_k\mathcal{P}_G^k
   \nonumber\\
&& \textstyle+\frac{i}{2}
\sum_kG^k_0\mathcal{P}^{12+}\sum_l(-1)^l\mathcal{P}_G^l
 - {L}^{12+}\mathcal{P}_G^1\mathcal{P}_G^2- \sum_{l}L^{1l}\mathcal{P}^{l2+} \mathcal{P}^{11+}\nonumber\\
&& \textstyle + \sum_{l}L^{l2}\mathcal{P}^{1l+}\mathcal{P}^{22+}-
\sum_{l}(-1)^lL^{ll}\mathcal{P}^{ll+} \mathcal{P}^{12+}\Bigr\}
\nonumber\\
&& \textstyle  r^2\eta_0\eta^+_i\eta_j\Bigl\{ \frac{i}{2}
\sum_{l}(-1)^lG^l_0\sum_k\mathcal{P}_G^k\mathcal{P}\delta^{1j}
\delta^{2i}  + 2(L^{22+}\mathcal{P}^{22}-L^{11}\mathcal{P}^{11+})
\mathcal{P}\delta^{1j} \delta^{2i} \nonumber\\
&&
-2T\mathcal{P}^{11}\mathcal{P}^{22+}\delta^{1i}\delta^{2j}+\textstyle\frac{1}{2}\varepsilon^{\{1j}\delta^{2\}i}\Bigl\{
iT\mathcal{P}^+\sum_k\mathcal{P}_G^k - 4i L^{12}\mathcal{P}^{12+}
\sum_{l}(-1)^l\mathcal{P}_G^l\Bigr\}
\nonumber\\
&&
   -T\mathcal{P}_G^1\mathcal{P}_G^2\delta^{2i} \delta^{1j}
+2\varepsilon^{\{1j}\delta^{2\}i}\Bigl\{(
T^+\mathcal{P}^{12}-L^{12}\mathcal{P}^+)\mathcal{P}^{11+}  +(
L^{12}\mathcal{P}-T\mathcal{P}^{12})\mathcal{P}^{22+}
\Bigr\}\nonumber\\
&& +
\textstyle2\sum_{l}(-1)^lG^l_0\mathcal{P}^{12+}(\mathcal{P}^{11}
\delta^{1i}\delta^{2j}- \mathcal{P}^{22}\delta^{2i}\delta^{1j})
\nonumber\\
&& \textstyle-2i({L}^{22}\delta^{2i}\delta^{1j}-
{L}^{11}\delta^{1i}\delta^{2j})\mathcal{P}^{12+}
\sum_{l}(-1)^l\mathcal{P}_G^l
  \Bigr\} + h. c.\end{eqnarray}
The property of the BRST operator $Q'$ to be Hermitian is defined
by the rule
\begin{eqnarray}\label{HermQb}
  Q^{\prime +}K = K Q'\,,\ \mathrm{for} \ K = \hat{1} \otimes K' \otimes \hat{1}_{gh}\,,
    \end{eqnarray}
with unity operators in $\mathcal{H}$, $\mathcal{H}_{gh}$, and
the operator $K'$ providing the Hermiticity of the additional parts
$o'_I$ in $\mathcal{H}'$, as in Refs.~\cite{flatfermmix,
adsfermBKR}. We first note that the BRST operator for the
superalgebra $\mathcal{A}_c$ is treated in the same way, with
the above nontrivial Jacobi identities, whose r.h.s. is extended
by a fermionic operator. Second, in the case $i=j=1$, (\ref{Q'nlb2})
yields a BRST operator for the converted algebra $\mathcal{A}_{bc}$
in the case of totally-symmetric bosonic HS fields \cite{0206027}.

\vspace{2ex} {\center\subsubsection{LAGRANGIAN
FORMULATIONS}}\label{lagrform} \vspace{1ex}

A covariant extraction of $G^i_0 = g_0^i +
g_0^{\prime i}(h^i)$ from
$\{\tilde{O}_I\}$, in order to pass to that part of the converted
1st-class constraints $\{{O}_\alpha\}$ which corresponds only to
equations (\ref{Eq-0b}), (\ref{Eq-1b}), i.e., the constraints
$\{L_0,L^i,L^{ij},T\}$, is based on a special representation for
$\mathcal{H}_{gh}$ in $\mathcal{H}_{tot} =
\mathcal{H}\otimes\mathcal{H}'\otimes\mathcal{H}_{gh}$, such that
the operators $(\eta_i, \eta_{ij},\eta, \mathcal{ P}_0, \mathcal{P}_i,
\mathcal{P}_{ij}, \mathcal{P},\mathcal{P}^{i}_G)$ should annihilate
the vacuum $|0\rangle$, as well as on the elimination from ${Q}'$
of the terms proportional to $\mathcal{P}^i_G, \eta^i_G,
\mathcal{K}^i = (\sigma^i+h^i)$, as in \cite{flatfermmix, 0707.2181},
\begin{eqnarray}
{Q}'  =  {Q} + \eta^i_G\mathcal{K}^i + \mathcal{B}^i
\mathcal{P}^i_G,\ \mathcal{K}^i =  G^i_0 + \bigl(
\textstyle\sum_{j}(1+\delta_{ij})\eta^{ij+}\mathcal{P}_{ij}
+(-1)^i\eta^+\mathcal{P} + h.c.\bigr). \label{decomposQ'}
\end{eqnarray}
The same applies to a scalar physical vector $|\chi\rangle \in
\mathcal{H}_{tot}$, $|\chi\rangle$=$|\Phi\rangle +
|\Phi_A\rangle$, $|\Phi_A\rangle_{\{(a, a^+)_\mathbf{a} =
\mathcal{C} = \mathcal{P} = 0\}}$ = $0$, with $|\Phi\rangle$ given
by (\ref{PhysState}) and with the use of the BFV--BRST equation
$\tilde{Q}'|\chi\rangle = 0$ that determines physical states,
\begin{equation}
\label{Qchi}  \tilde{Q}|\chi\rangle = 0, \quad
(\sigma^i+h^i)|\chi\rangle=0, \quad \left(\varepsilon,
{gh}\right)(|\chi\rangle)=(0,0).
\end{equation}
Notice that the second equations must take place in the entire
$\mathcal{H}_{tot}$, thus determining the spectrum of spin values
for $|\chi\rangle$ and the corresponding proper eigenvectors,
\begin{eqnarray}
h^i&=&\textstyle -\bigl( s_i + \frac{d - 5 }{2} -2\delta^{i2}
\bigr)\,, \label{hchin1n2} \qquad (s_1,s_2) \in (\mathbb{Z},
\mathbb{N}_0), \ |\chi\rangle_{(s_1,s_2)},
\end{eqnarray}
whereas the first equation is valid only in the subspace of
$\mathcal{H}_{tot}$ with the zero ghost number.

Because of the commutativity of $\sigma^i$ with $Q$, the latter,
being subject to the substitution $h^i\to -\bigl( s_i + \frac{d -
5 }{2} -2\delta^{i2} \bigr)$, i.e., $Q \to Q_{(s_1,s_2)}$, is
nilpotent in each of the subspaces $H_{tot{}(s_1,s_2)}$ whose
vectors obey Eqs.~(\ref{Qchi}), (\ref{hchin1n2}). Thus, the
equations of motion are in a one-to-one correspondence with
Eqs.~(\ref{Eq-0b}), (\ref{Eq-1b}), whereas the sequence of
reducible gauge transformations has the form
\begin{eqnarray}
Q_{(s_1,s_2)}|\chi^0\rangle_{(s_1,s_2)}=0,  \ \delta|\chi^{l}
\rangle_{(s_1,s_2)} =Q_{(s_1,s_2)}|\chi^{1+1}\rangle_{(s_1,s_2)},
\ l = 0,...,6,\label{LEoM}
\end{eqnarray}
for $|\chi^0\rangle\equiv |\chi\rangle$, and can be obtained from
the Lagrangian action
\begin{eqnarray}
{\cal S}_{n_1,n_2} = \int d \eta_0 \; {}_{(s_1,s_2)}\langle \chi^0
|K_{(s_1,s_2)} Q_{(s_1,s_2)}| \chi^0 \rangle_{(s_1,s_2)} \
K_{(s_1,s_2)}=K\vert_{h^i\to   -\bigl( s_i + \frac{d - 5 }{2}
-2\delta^{i2} \bigr)}, \label{S}
\end{eqnarray}
where the standard $\varepsilon$-even scalar product in
$\mathcal{H}_{tot}$ is assumed.

The corresponding LF of a bosonic  field with a specific value
of spin $\mathbf{s}$ subject to $Y(s_1,s_2)$ is an unconstrained
reducible gauge theory of $L = 5$-th stage of reducibility.

\vspace{2ex}
{\center\subsubsection{CONCLUSIONS}}\label{Conclusions}
\vspace{1ex}

We have found and studied the properties of nonlinear operator
algebras and superalgebras underlying integer and half-integer
HS fields in an AdS$_d$ space that are subject to a Young tableaux
with two rows. To construct a BFV--BRST operator, whose special
cohomology in the corresponding Hilbert space with a vanishing
ghost number should coincide with the space of solutions for
the equations that determine (spin-)tensors of the AdS-group
irreducible representation with a given mass and generalized spin,
we formulate a proposition that determines the form of algebraic
relations both for the parts additional to the operators of
the initial polynomial superalgebra with given relations and for
the superalgebra $\mathcal{A}_{c}$ of additively converted
operators. We have briefly shown, first, the solvability of the
algorithm of constructing the Verma module for the superalgebra of
the additional parts, and, second, have described a way to realize
it in terms of a formal power series in the corresponding
Fock-space variables, which completes our procedure of converting
the subset with 2nd-class constraints into that of converted
1st-class ones. We have obtained an exact BFV--BRST operator
having terms of at most 3rd degree in ghost coordinates
for a general quadratic operator superalgebra having nontrivial
3rd-order structure functions and relations that resolve the Jacobi
identities for $\mathcal{A}_c$ only. Following this prescription,
we have found an exact BRST operator (\ref{Q'b}) for a nonlinear
algebra $A_{bc}$ of converted operators. Finally, on a basis of
the resulting BRST operator, we have developed a gauge-invariant
approach to an unconstrained LF, given by Eqs.~(\ref{LEoM}), (\ref{S}),
which describes the dynamics of free bosonic HS fields in an AdS$_d$
space with an index symmetry corresponding to a two-row Young tableaux.
We must draw the attention of the reader to some problems that have
remained not entirely solved: an explicit form of the Verma module
for the (super)algebra in question, the deduction of a BRST
operator for a converted superalgebra with fermionic HS fields, its
application to a construction of the corresponding LF, and a more
detailed development of the constrained~LF.

The author is grateful to I.A. Batalin for the comments and
indication the typos, to  P.Yu. Moshin for useful discussions and
thanks the organizers  of the ``XXVII International Colloquium on
Group Theoretical Methods in Physics'', Yerevan, Armenia, August
2008, for support and hospitality. The work was supported by an
RFBR grant, project No. 08-02-08602.

\end{document}